\documentclass[11pt,a4paper]{article}
\usepackage{amsmath}
\usepackage{algorithm}
\usepackage{algpseudocode}
\usepackage{breakcites}
\usepackage{natbib}
\usepackage{graphicx}
\usepackage{hyperref}
\providecommand{\keywords}[1]
{
  \small
  \textbf{\textit{Keywords:}} #1
}
\begin{document}
%
\title{\bf Stochastic Predictive Analytics for Stocks
           in the Newsvendor Problem}

\author{Pedro A.\ Pury\footnote{ORCID
\href{https://orcid.org/0000-0003-2229-0590}{0000-0003-2229-0590}
}
\\
Facultad de Matem\'atica, Astronom\'\i a,
F\'\i sica y Computaci\'on \\
Universidad Nacional de C\'ordoba \\
Ciudad Universitaria, X5000HUA C\'ordoba, Argentina \\
\texttt{pedro.pury@unc.edu.ar}}

\date{\today}
%
\maketitle
%
\begin{abstract}
This work addresses a key challenge in inventory management
by developing a stochastic model that describes the dynamic
distribution of inventory stock over time without assuming
a specific demand distribution.
Our model provides a flexible and applicable solution
for situations with limited historical data and short-term
predictions, making it well-suited for the Newsvendor problem.
We evaluate our model's performance using real-world data
from a large electronic marketplace, demonstrating
its effectiveness in a practical forecasting scenario.
\end{abstract}
%


\vspace{0.2cm}

\section{Introduction}
\label{sec:Intro}

The Newsvendor problem is a fundamental model in inventory
management~\citep{Rossi21} that accommodates both
known~\citep{DKW52a} and unknown~\citep{DKW52b} demand distributions.
Since its inception~\citep{Edg1888}, it has been widely applied
in inventory control and policy-making~\citep{AHM51},
as well as various real-world situations~\citep{Choi12,CCCW16}.
Its simplicity stems from considering a single product for sale,
for which the optimal initial stock level must be determined
to satisfy forecasted demand over a given period
without restocking.
The interplay among purchasing cost, selling price, and stock
ordered at the beginning of the period determines the inventory
management policies~\citep{Whi52,Ros54,PD99}.
The model has been extensively studied
for single stock-keeping units (SKUs).

Electronic marketplaces introduce an extra complication
to the problem, as they need to manage a large number
of SKUs at distribution centers alongside highly variable
demand received through electronic platforms.
This situation requires dealing with short demand histories
to forecast in a short time horizon and designing forecasting
methods that are easily upgradable with new data and have low
computational costs for inference.

Usually, forecasting methods in this context are intended
for point forecasts of stock value.
However, for practical and fundamental reasons, it is highly
convenient to know the distribution of stock over time,
given the initial level.
This is a valuable tool for planning and decision-making
in a wide variety of inventory situations.
Without replenishment, the Newsvendor's stock over time follows
a non-increasing discrete-time series.
Stochastically non-increasing Markov chains have been studied
for a long time~\citep{Dal68}, and in recent years, particular
attention has been given to scaling limits~\citep{HM11}.
However, explicit solutions in specific cases,
such as the Newsvendor problem, are currently lacking.

This study contributes to the existing literature by developing
a stochastic model that describes the dynamic distribution
of inventory stock over time for a single stock-keeping unit (SKU)
within a single-period planning horizon.
Drawing inspiration from Croston's method~\citep{Cro72},
which decomposes demand into sizes and intervals for forecasting,
we construct the time distribution of stock by analyzing
stochastic sales dynamics and demand distribution separately.
This approach yields a time distribution that is agnostic
to the demand distribution considered.

We further innovate by deriving specific instances of the model
for various parametric and non-parametric demand distributions,
showcasing its flexibility and applicability.
Our approach is well-suited for the Newsvendor problem,
particularly in situations where historical data are limited
and short-term predictions are required.
By understanding the distribution of stock over time,
inventory managers can make more informed decisions about
inventory optimization, supply chain management,
and resource allocation, ultimately leading to reduced costs
and improved efficiency through better restocking policies.

We evaluate our model's performance in a real-world forecasting
scenario using actual data from an electronic marketplace
covering a wide range of SKUs.
We demonstrate the effectiveness of our approach in the practical
forecasting scenario of predicting stockouts, providing valuable
insights for both academics and decision-making practitioners
in retail management.
Thus, the work is intended to present a useful forecasting case,
but the generic nature of the methodology is emphasized.

This paper is organized as follows:
Sections~\ref{sec:Model},~\ref{Sec:Demand}
and~\ref{Sec:Evaluation} cover Materials and Methods.
In Section~\ref{sec:Model}, we introduce our stochastic model,
derive the solution for the time-dependent stock distributions,
and analyze frustrated sales due to stockouts.
In Section~\ref{Sec:Demand}, we complement the general stochastic
framework by considering several demand distributions, ranging from
a simple count of daily sales frequencies to various parametric
demand distributions, including deterministic, Poisson, binomial,
and negative binomial models.
Section~\ref{Sec:Evaluation} applies these theoretical results
to stockout forecasting, encompassing evaluation metrics,
establishment of a baseline value, description of the dataset
used in the experiments, and proposal of a benchmark for evaluations.
Furthermore, we describe implementation methods
for the naive frequentist non-parametric approach to demand modeling,
as well as parameter estimation techniques for the distributions
mentioned in the previous section.
The results of stockout forecasting are presented
in Section~\ref{sec:FR}, followed by conclusions
in Section~\ref{sec:fin}.

\section{Stochastic Model for Stocks}
\label{sec:Model}

In the absence of replacements or repositions, we can model
the daily stock of a specific SKU as a monotonic non-increasing
discrete-time series of non-negative integer values.
The object of our interest is the probability, $P(n,k)$, of having
in stock $n$ items on day $k$, given an initial supply of $m>1$ items,
given the initial condition
\begin{equation}
P(n,k=0) = \delta_{n,m} \,,
\label{initial}
\end{equation}
where $\delta_{n,m}$ is the Kronecker delta,
and with the condition of normalization
\begin{equation}
\sum_{n=0}^m P(n,k) = 1, \;\forall k \geq 0.
\label{normal}
\end{equation}
The stochastic shopper's demand is characterized by the probability
per unit time, $\alpha_{l}$, which represents the probability
of selling $l$ units of the product within a day.
In this case, the condition of normalization results
\begin{equation}
\displaystyle\sum_{\ell=0}^{\infty} \alpha_\ell = 1 \,.
\label{normala}
\end{equation}
We further define the probability per unit time,
$\beta_n = \displaystyle\sum_{\ell=n}^{\infty} \alpha_\ell$,
of selling at least $n$ items within a day.
Specifically, $\beta_n$ is associated with the shoppers'
intention to independently purchase at least $n$ items on that day,
irrespective of the current stock level.
It follows from Eq.~(\ref{normala}) that $\beta_0=1$.

\subsection{Markov model}
\label{sec:Markov}

We assume that $\alpha_{\ell}$ is independent of the current stock
level. Furthermore, we assume that the shopper's purchasing behavior
remains consistent throughout the observation period,
allowing us to consider as $\alpha_{\ell}$ constant
within this window.
The assumption that $\alpha_{\ell}$ remains constant over time
is a simplification, as day-of-week seasonality may be present
for some products. When dealing with the short length of demand
series, more complex models may not be feasible, but the potential
limitations of this approach should be recognized.

Under these assumptions, we have a Markov process and we can write
the following recurrence equations for the probabilities of interest
\begin{eqnarray}
P(0,k) &=& \sum_{n=0}^m \beta_n \,P(n,k-1) \,,
\label{ME0} \\
P(n,k) &=& \sum_{\ell=n}^m \alpha_{\ell-n} \,P(\ell,k-1),
\mbox{ for } 1 \leq n \leq m \,.
\label{MEn}
\end{eqnarray}
In the Newsvendor context, $P(0,k)$ represents the first-passage
or absorption time distribution to the stock level $0$,
which corresponds to the stockout time distribution.
Using these recurrence equations, one can easily prove by complete
induction the normalization given by Eq.~(\ref{normal}),
where the base case follows directly from Eq.~(\ref{initial}).
Alternatively, we can assume that the normalization condition
is satisfied and prove by complete induction the recurrence
Eq.~(\ref{ME0}) from Eq.~(\ref{MEn}), where the base case
also follows directly from the initial condition.

From Eqs.~(\ref{MEn}) and~(\ref{initial}), we can write
\begin{equation}
P(m,k) = \alpha_0 \,P(m,k-1) = \alpha_0^k \,P(m,k=0) = \alpha_0^k \,.
\end{equation}
The last equation implies that $P(m,k=0) = 1$ and, given that
$0 < \alpha_0 < 1$, it directly shows that $P(m,k)$ is a monotonic
decreasing function of k.
On the other hand, from Eq.~(\ref{ME0}), using that $\beta_0 = 1$
results
\begin{equation}
P(0,k) - P(0,k-1) = \sum_{n=1}^m \beta_n \,P(n,k-1) > 0 \,.
\label{aux1}
\end{equation}
Therefore, $P(0,k)$ is a monotonic increasing function of $k$
with $P(0,k=0) = 0$.

Using Eq.~(\ref{normala}), we can recast $\beta_n$ as
\begin{equation}
\beta_n = 1 - \displaystyle\sum_{j=0}^{n-1} \alpha_j \,.
\label{beta}
\end{equation}
Then, substituting the last equation in Eq.~(\ref{aux1}), we obtain
\begin{equation}
\sum_{n=1}^m \sum_{j=0}^{n-1} \alpha_j \,P(n,k-1) = 1 - P(0,k) \,.
\label{aux2}
\end{equation}
%

\subsection{Recursive Solutions}
\label{Sec:RS}

Using the initial condition~(\ref{initial}), from Eq~(\ref{ME0})
we inmmediately obtain $P(0,1) = \beta_m$, and $\forall k \geq 2$
we can also write
\begin{equation}
P(0,k) = \beta_0^{k-1} \beta_m
+ \sum_{j=1}^{k-1} \beta_0^{k-1-j} \sum_{n=1}^m \beta_n P(n,j) \,.
\label{P(0,k)}
\end{equation}
Trivially, the solution~(\ref{P(0,k)}) can be rewritten as
Eq.~(\ref{ME0}).

Again, using the initial condition~(\ref{initial}), from Eq~(\ref{MEn})
we immediately obtain $P(n,1) = \alpha_{m-n}$, for $1 \leq n \leq m$,
and $\forall k>0$, we can also write
\begin{equation}
P(n,k) = \sum_{\ell_{k-1}=n}^m \alpha_{\ell_{k-1}-n}
\sum_{\ell_{k-2}=\ell_{k-1}}^m \alpha_{\ell_{k-2}-\ell_{k-1}}
\cdots
\sum_{\ell_1=\ell_2}^m \alpha_{\ell_1-\ell_2} \,\alpha_{m-\ell_1}
\,.
\label{P(n,k)}
\end{equation}
It is direct to see that the solution~(\ref{P(n,k)})
satisfies Eq.~(\ref{MEn}). Additionally, we can easily observe
that $P(m,k) = \alpha_0^k$. Moreover, given that $0 < \alpha_j < 1$,
we also obtain that $P(n,k) \rightarrow 0$
for $k \rightarrow \infty$ if $n \neq 0$.
Then, from Eq.~(\ref{normal}) results $P(0,k) \rightarrow 1$
for $k \rightarrow \infty$.

Equations~(\ref{P(0,k)}) and~(\ref{P(n,k)}) provide a closed-form
solution for the relevant probabilities; however, they rely
on the assumed models for the probability $\alpha_\ell$
that enables the fulfillment of all involved calculations.
For their part, the recurrence relations presented in
Eqs.~(\ref{ME0}) and~(\ref{MEn}), along with the initial
conditions $P(0,1) = \beta_m$ and $P(n,1) = \alpha_{m-n}$,
allow for efficient numerical computation of $P(0,k)$.

Letting $Y[k]$ denote the probability $P(0,k)$,
we aim to calculate this probability recursively over
a time window of $N$ days ($1 \leq k \leq N$).
To achieve this, we utilize the Algorithm~\ref{algo},
which leverages previously computed probabilities for $k-1$
on each iteration. Additionally, we assume the vector
$\alpha[\ell]$ is known.
\begin{algorithm}[hbt!]
\begin{algorithmic}
     \State $Y \gets \mbox{repeat}(0,N)$
     \State $P \gets \mbox{repeat}(0,m)$
     \For {k in (1:N)}
          \If {k == 1}
               \State $Y[1] \gets \beta[m]$
               \For {n in (1:m)}
                    \State $P[n] \gets \alpha[m-n]$
               \EndFor
          \Else
               \State $P\mbox{old} \gets P$
               \State $Y[k] \gets \beta[0] * Y[k-1]$
               \For {n in (1:m)}
                    \State $Y[k] \gets  Y[k] + \beta[n] * P\mbox{old}[n]$
                    \State $P[n] \gets 0$
                    \For {$\ell$ in (n:m)}
                         \State $P[n] \gets P[n] + \alpha[\ell-n] * P\mbox{old}[\ell]$
                    \EndFor
               \EndFor
          \EndIf
     \EndFor
\end{algorithmic}
\caption{Recursive solution for $P(0,k) \equiv Y[k]$
for $k \in [1,N]$ with initial stock of $m$ items.}
\label{algo}
\end{algorithm}

\subsection{Frustrated Sales}
\label{Sec:FS}

Online retailer's stockout policy has significant impact on
consumers' purchase behavior and choice decisions~\citep{BCG06}.
Estimating the number of potential sales lost in a marketplace
is essential for understanding the impact of stockouts
on a retailer's performance and customer satisfaction~\citep{THP24}.
\citet{Kar57} examined multi-period inventory management
for a commodity, utilizing a queueing model to derive
the stationary-state probability $p(n,m)$, or fraction of time,
of having $n$ units in in the long-run, given that $m$ is the total
fixed inventory, under assumptions of Poisson demand and exponential
resupply times.
The fraction of lost sales with respect to the total demand
is $p(0,m)$. This fraction represents the proportion of units
that are not sold due to stockouts, and it is a key performance
metric for inventory management systems.
In the one-period case without stock replenishment,
the process is transient to the stock level $0$,
and the probability of lost sales at time $k$
is directly given by $P(0,k)$.

Lost sales are associated with inventory depletion.
However, limited inventory also poses the risk of unsatisfied
customers. In some instances, even if stock is available
at the beginning of the day, customers who saw the product
advertised early may still be left without their preferred
products, or at least not in the desired quantity,
due to insufficient stock later on.
In this context, we refer to {\em frustrated sales}
as the number of potential sales that exceed available stock.
Despite its significance, this phenomenon has received limited
attention in the literature.
Notably, frustrated sales highlight that even with non-zero
inventory levels, loss of sales can occur, necessitating
adjustments in stock management to mitigate their impact.
Although data on frustrated sales are censored,
they can be estimated from our stock distribution.

In our framework, for a particular day $k \geq 1$, the probability
of frustrated sales is given by
\begin{equation}
P_F(k) = \sum_{n=1}^m \beta_{n+1} \,P(n,k-1) \,,
\label{PF}
\end{equation}
with $P_F(0) = 0$. Clearly, $P_F(k) \geq 0$ for all $k$.
From Eq.~(\ref{beta}) we can recast the last expression in the form
\begin{equation}
P_F(k) = 1 - P(0,k-1) -
\sum_{n=1}^m \sum_{j=0}^{\bf n} \alpha_j \,P(n,k-1) \,,
\end{equation}
and using Eq.(\ref{aux2}), for $k>1$ we obtain
\begin{equation}
P_F(k) = P(0,k) - P(0,k-1) - \sum_{n=1}^m \alpha_n \,P(n,k-1) \,.
\label{PFk}
\end{equation}
Given the behaviors discussed in Sec.~\ref{sec:Markov},
we have that $P_F(k) \rightarrow 0$ for $k \rightarrow \infty$.
In contrast, the probability of lost sales behaves as
$P(0,k) \rightarrow 1$ for $k \rightarrow \infty$.
This behavior may seem counterintuitive, but it is a direct
consequence of the fact that without stock, there can be no
frustrated sales. Frustrated sales only occur when there is stock,
but it is insufficient to satisfy the demand.
In consequence, the probability of frustrated sales has a maximum
as function of $k$.
Moreover, if for a particular day $k_0 > 1$ we have had a stockout
the previous day, $P(0,k_0-1)=1$, then $P(n,k_0-1)=0$
for $1 \leq n \leq m$, $P(0,k_0)=1$, and consequently, $P_F(k_0)=0$.

\section{Demand distributions}
\label{Sec:Demand}

Typically, extrapolation methods for forecasting future values
of a time series rely on past values, assuming that past trends
and patterns will persist into the future, with varying
consideration of demand correlations~\citep{AXYY16}.
This assumption is fundamental to autoregressive models,
which rely on the concept of stationarity. A time series
is considered stationary if its properties remain invariant
to the observation time.

Equations (\ref{P(0,k)}) and (\ref{P(n,k)}) are agnostic
to the demand distributions. Hence, under the above assumption,
we can employ various demand distributions to complete
the stochastic model for inventory systems presented
in Sec.~\ref{sec:Model}.
In this way, we can model demand using a straightforward approach,
such as the frequentist distribution or parametric discrete
distributions~\citep{JKK05}
(e.g., Poisson~\citep{Whi52,Kar57,Hil96,Bij14,RPAH14},
binomial~\citep{Hil96, RPAH14},
and negative binomial~\citep{Bij14,AS96,SFGJ20}),
which are empirically grounded in inventory contexts,
or a simple deterministic case with a constant demand rate.

\subsection{Naive frequentist distribution}
\label{Sec:naive}

The most straightforward approach is the frequentist distribution,
which is derived from historical data by calculating the frequencies
of daily sales numbers over a specific period immediately preceding
the forecasting time window.
This very simple density forecasting method, the ``climatological''
approach, is very practical and applicable in real-world situations
where theoretical models for the probability $\alpha_\ell$
may be unknown or difficult to obtain.
The recursive calculation proposed in Sec.~\ref{Sec:RS},
adapts perfectly to deal with this case.

\subsection{Deterministic distribution}
\label{Sec:deter}

As a first parametric demand distribution, we consider
a straightforward scenario where it is intuitive to forecast
outcomes and assess model consistency. Specifically, we assume
that a constant number of items of the product are sold every day
in a systematic manner.

If $h \geq 1$ is the exact number of sales per day,
we have $\alpha_\ell = \delta_{\ell,h}$. Using that
\begin{equation}
\displaystyle\sum_{\ell_j=\ell_{j+1}}^m \delta_{\ell_j,\ell_{j+1}+h}
\,\delta_{\ell_j,m-(k-1)h} = \delta_{\ell_{j+1},m-kh} \,,
\label{deltas}
\end{equation}
we obtain from Eq.~(\ref{P(n,k)}) for $k \geq 1$
and $1 \leq n \leq m$,
\begin{equation}
P(n,k) = \delta_{n,m-kh} \,.
\label{deterPn}
\end{equation}
It is immediate to see that the initial condition~(\ref{initial})
is obtained and that $P(m,k) = \delta_{k,0}$, as expected.
As can be seen in the Appendix~\ref{Appendix}, the last equation
can be proven by complete induction. In this way,
\begin{equation}
\sum_{n=1}^m P(n,k) = \left\{
\begin{array}{ll}
1, & 1 \leq m - k \,h \leq m \,, \\
0, & m - k \,h < 1 \,.
\end{array}
\right.
\end{equation}
Then,
\begin{equation}
P(0,k) = \left\{
\begin{array}{ll}
0, & k \,h \leq m-1 \,, \\
1, & k \,h \geq m \,,
\end{array}
\right.
\label{deterP0}
\end{equation}
and
\begin{equation}
\sum_{n=1}^m \alpha_n \,P(n,k-1) =
\sum_{n=1}^m \delta_{n,h} \delta_{n,m-(k-1)h} = \delta_{m,kh} \,,
\end{equation}
that is, the summation is not null only if $m$ is a multiple
of $h$. Also, we get
\begin{equation}
P(0,k) - P(0,k-1) = \left\{
\begin{array}{ll}
0, & k \,h \leq m-1 \mbox{ or } k \,h \geq m+h \,, \\
1, & m \leq k \,h \leq m+h-1 \,.
\end{array}
\right.
\end{equation}
Thus, from Eq.~(\ref{PFk}), the probability of frustrated sales
results
\begin{equation}
P_F(k) = \left\{
\begin{array}{ll}
0, & k \,h \leq m \mbox{ or } k \,h \geq m+h \,, \\
1, & m+1 \leq k \,h \leq m+h-1 \,.
\end{array}
\right.
\end{equation}
In this manner, if $m = p \,h$ with $p$ a positive integer,
we get $P_F(k) = 0$ for all $k \geq 1$, as it is expected.

\subsection{Poisson distribution}
\label{Sec:Poisson}

Sales follow a Poisson process if the numbers of sales
in non-overlapping time intervals are mutually independent,
the probability of a sale within a given time interval is constant,
and the probability of two or more sales in a small time interval
is negligible. Additionally, the probability of observing
a single sale in a small time interval is approximately
proportional to the interval's length~(\cite{BS21}).
Under these assumptions, the Poisson probability distribution
with parameter $\lambda$ is given by
\begin{equation}
\alpha_\ell = \frac{\lambda^\ell}{\ell!} \, e^{-\lambda} \,,
\label{Poisson}
\end{equation}
where $\lambda$ represents the average daily sales.

Using the binomial expansion
\begin{equation}
\frac{k^N}{N!} = \sum_{j=0}^N \frac{(k-1)^{N-j}}{j! \,(N-j)!} \,,
\label{binomial}
\end{equation}
we obtain from Eq.~(\ref{P(n,k)}) for $k \geq 1$
and $1 \leq n \leq m$,
\begin{equation}
P(n,k) = e^{-k \,\lambda}
\frac{(k \,\lambda)^{m-n}}{(m-n)!} \,.
\label{PoissonPn}
\end{equation}
The last equation satisfies $P(n,k=0) = \delta_{n m}$
and $P(m,k) = \alpha_0^k$.
Eq.~(\ref{PoissonPn}) can be proved by complete induction.
The details can be seen in the Appendix~\ref{Appendix}.

From the condition of normalization we can calculate $P(0,k)$ as
\begin{equation}
P(0,k) = 1 - \displaystyle\sum_{n=1}^m P(n,k)
= 1 - e^{-k \,\lambda} \displaystyle\sum_{j=0}^{m-1}
\displaystyle\frac{(k \lambda)^j}{j!} \,.
\end{equation}
According to~\citet{AS65}, the last expression can be recast as
\begin{equation}
P(0,k) = 1 - \frac{\Gamma(m, k \,\lambda)}{(m-1)!} \,,
\label{PoissonP0}
\end{equation}
where $\Gamma(a,x)$ is the upper incomplete gamma function.
Given that $\Gamma(m,0)=(m-1)!$, we reobtain $P(0,k=0) = 0$ and from
$\Gamma(m, k \,\lambda) \rightarrow 0$ for $k \rightarrow \infty$,
we get $P(0,k\rightarrow \infty) = 1$.

Using Eqs.~(\ref{Poisson}--\ref{PoissonPn}) we can write
\begin{equation}
\displaystyle\sum_{n=1}^m \alpha_n \,P(n,k-1) =
e^{-k \,\lambda} \,\lambda^m \left(
\displaystyle\frac{k^m}{m!} - \displaystyle\frac{(k-1)^m}{m!}
\right)
\end{equation}
Thus, form Eq.~(\ref{PFk}) results
\begin{equation}
P_F(k) = \frac{\Gamma(m,(k-1) \,\lambda)
              -\Gamma(m,k \,\lambda)}{(m-1)!}
- e^{-k \,\lambda} \frac{(k \,\lambda)^m - ((k-1) \,\lambda)^m}{m!}
\end{equation}
and using the recurrence relation
$\Gamma(m+1,x) = m \,\Gamma(m,x) + x^m \,e^{-x}$,
we finally obtain
\begin{equation}
P_F(k) = e^{-k \,\lambda} \frac{((k-1) \,\lambda)^m}{m!}
       + \frac{\Gamma(m,(k-1) \,\lambda)}{(m-1)!}
       - \frac{\Gamma(m+1,k \,\lambda)}{m!} \,.
\label{PoissonPF}
\end{equation}
%

\subsection{Binomial distribution}
\label{Sec:Binomial}

The Binomial probability distribution with parameters $(C, \,p)$
is given by
\begin{equation}
\alpha_\ell = \binom{C}{\ell} \, p^\ell \,q^{C-\ell} \,,
\label{Binomial}
\end{equation}
where $C$, in our context, is the number of independent customers
each having a probability $p$ of purchasing a single unit within
each day and $q = 1-p$. For our problem $C \geq m$.

Using the Vandermonde identity~\citep{AS65}
\begin{equation}
\binom{M+N}{r} = \sum_{\ell=0}^r \binom{M}{\ell} \,\binom{N}{r-\ell} \,,
\label{Vandermonde}
\end{equation}
we obtain from Eq.~(\ref{P(n,k)}) for $k \geq 1$
and $1 \leq n \leq m$,
\begin{equation}
P(n,k) = \binom{kC}{m-n} \,p^{m-n} \,q^{kC-(m-n)} \,.
\label{BinomialPn}
\end{equation}
The last equation satisfies $P(m,k) = q ^{kC} = \alpha_0^k$,
given that $\alpha_0 = q^C$.
Eq.~(\ref{BinomialPn}) can be proved by complete induction.
The details are relegated to the Appendix~\ref{Appendix}.

From the condition of normalization we can calculate $P(0,k)$
and results
\begin{equation}
P(0,k) =
\displaystyle\sum_{j=m}^{kC}
\binom{kC}{j} \,p^{j} \,q^{kC-j} \,.
\end{equation}
This expression can also be written as~\citep{AS65}
\begin{equation}
P(0,k) = I_p(m,kC-m+1) \,,
\label{BinomialP0}
\end{equation}
for $kC-m+1 \geq 0$, where $I_x(a,b)$ is the regularized incomplete
beta function. If $kC-m+1 < 0$ then $P(0,k)=0$.

Using Eqs.~(\ref{Binomial}), (\ref{BinomialPn})
and~(\ref{Vandermonde}) we can write
\begin{equation}
\displaystyle\sum_{n=1}^m \alpha_n \,P(n,k-1) =
\left(
\displaystyle\binom{kC}{m} - \binom{(k-1)C}{m}
\right) \,
p^m \,q^{kC-m}
\end{equation}
Thus, from Eq.~(\ref{PFk}) results
\begin{equation}
P_F(k) = I_p(m,kC-m+1) - I_p(m,(k-1)C-m+1)
- \left( \binom{kC}{m} - \binom{(k-1)C}{m} \right) \,
p^m \,q^{kC-m} \,,
\end{equation}
and using the relation
\begin{equation}
P(0,k) = \displaystyle\sum_{j=m+1}^{kC}
\binom{kC}{j} \,p^{j} \,q^{kC-j} + \binom{kC}{m} \,p^m \,q^{kC-m} \,,
\end{equation}
we finally obtain
\begin{equation}
P_F(k) = I_p(m+1,kC-m) - I_p(m,(k-1)C-m+1)
+ \binom{(k-1)C}{m} p^m \,q^{kC-m} \,.
\label{BinomialPF}
\end{equation}
%

\subsection{Negative Binomial distribution}
\label{Sec:NBinomial}

There are two distinct ways in which this distribution may arise
as a demand distribution.
Firstly, it can emerge as a compound distribution, where the number
of demand incidences follows a Poisson distribution,
and the demand sizes follow a logarithmic distribution.
Alternatively, it can arise as a mixture distribution, where demand
is Poisson distributed, and the mean demand varies over time
according to a gamma distribution~(\cite{BS21}).

The Negative Binomial probability distribution with parameters
$(r, \,p)$ is given by
\begin{equation}
\alpha_\ell = \binom{r-1+\ell}{\ell} \, p^r \,q^{\ell} \,,
\label{NBinomial}
\end{equation}
where $r \geq \ell$.
In our context, $r$ represents the number of daily visits
to a physical store or electronic marketplace website
where the SKU is displayed without successful transactions,
until $\ell$ sales are achieved.
Meanwhile, $p$ represents the probability of a shopper visiting
that marketplace without making a purchase, while $q=1-p$ symbolizes
the probability of successfully selling the SKU on a single visit.

Using a generalization of Vandermonde's convolution~\citep{Gou56},
\begin{equation}
\sum_{\ell=0}^n \binom{\alpha+\ell}{\ell} \,\binom{\gamma+n-\ell}{n-\ell}
= \binom{\alpha+\gamma+1+n}{n}\,,
\label{Vandermonde2}
\end{equation}
we obtain from Eq.~(\ref{P(n,k)}) for $k \geq 1$
and $1 \leq n \leq m$,
\begin{equation}
P(n,k) = \binom{kr-1+m-n}{m-n} \,p^{kr} \,q^{m-n} \,.
\label{NBinomialPn}
\end{equation}
The last equation satisfies $P(m,k) = p^{kr} = \alpha_0^k$,
where $\alpha_0 = p^r$.
Eq.~(\ref{NBinomialPn}) can also be proved by induction.
The details can be found in the Appendix~\ref{Appendix}.

From the condition of normalization we can calculate $P(0,k)$ as
\begin{equation}
P(0,k) =
1 - \displaystyle\sum_{j=0}^{m-1}
\binom{kr-1+j}{j} \,p^{kr} \,q^{j} \,.
\end{equation}
The cumulative distribution function until $n$ of the negative
binomial distribution with parameters $(kr,p)$ is given by
$1-I_q(n+1,kr)$, where $I_x(a,b)$ is again
the regularized incomplete beta function.
Thus, we obtain
\begin{equation}
P(0,k) = I_q(m,kr) \,.
\label{NBinomialP0}
\end{equation}

Using Eqs.~(\ref{NBinomial}), (\ref{NBinomialPn})
and~(\ref{Vandermonde2}) we can write
\begin{equation}
\displaystyle\sum_{n=1}^m \alpha_n \,P(n,k-1) =
\left(
\displaystyle\binom{kr-1+m}{m} - \binom{(k-1)r-1+m}{m}
\right) \,
p^{kr} \,q^{m} \,.
\end{equation}
Thus, form Eq.~(\ref{PFk}) results
\begin{equation}
P_F(k) = I_q(m,kr) - I_q(m,(k-1)r) -
\left(
\displaystyle\binom{kr-1+m}{m} - \binom{(k-1)r-1+m}{m}
\right) \,
p^{kr} \,q^{m} \,,
\end{equation}
and using the relation
\begin{equation}
I_q(m+1,kr) = I_q(m,kr) - \binom{kr-1+m}{m} \,p^{kr} \,q^{m}\,,
\end{equation}
we finally obtain
\begin{equation}
P_F(k) = I_q(m+1,kr) - I_q(m,(k-1)r)
+ \binom{(k-1)r-1+m}{m} p^{kr} \,q^{m} \,.
\label{NBinomialPF}
\end{equation}
%

\section{Forecasting of Stockouts}
\label{Sec:Evaluation}

In this section, we evaluate our framework's performance
in a real-world forecasting scenario using actual data,
focusing on challenges with limited historical information.
Specifically, we examine problems involving very short
time series and short-term forecasting horizons.
In such cases, data scarcity typically hinders verification
of consistency conditions and effective set up of autoregressive
models, such as ARIMA (AutoRegressive Integrated Moving
Average)~\citep{Rossi21} or more complex variants.
However, these scenarios are of great importance in the context
of the Newsvendor problem in general.

Our stochastic model enables the computation of $P(n,k)$
for any initial stock level $m$, where $0 \leq n \leq m$,
and for discrete time $k \geq 0$.
One of the most compelling problems is forecasting
time series of marketplace sales, where we can specifically
compute the sold-out distribution over time, $P(0,k)$.
Stockouts are a critical issue, characterized by a complex
interplay between demand forecasting and inventory control.

Competitions and their released datasets have been enormously
important for learning how to improve forecasting accuracy
and compare methodologies.
The domain of the data may determine the characteristics
of the series, thus influencing the performances of the methods
being utilized~\citep{SKAM20}.
For our forecasting purposes, we will utilize the dataset released
by MercadoLibre~Inc.\ in 2021~\citep{meli2021}.

\subsection{Forecasting metric}
\label{Sec:RPS}

In this contribution, we focus on probabilistic forecasts
of uncertain outcomes, which necessitates a performance metric
that evaluates the entire probability distribution.
The concept of probabilistic forecasting was introduced
by~\citet{Bri50} in the context of weather forecasting.
Later, the concept was generalized by~\citet{Eps69},
who introduced the Ranked Probability Score (RPS),
extensively discussed by Murphy~\citep{Mur70,Mur71,GM92}
and~\citet{Fri83}.
RPS provides comprehensive insight into the distribution.
Unlike point estimation metrics such as Mean Absolute Error (MAE)
and Root Mean Squared Error (RMSE), which primarily assess
the accuracy of point forecasts.

RPS is a proper scoring rule that evaluates the quality
of density forecasts. It penalizes forecasts that are
over-confident or biased, thereby encouraging consideration
of distance in predictions~\citep{JNW09}.
Thus, it penalizes those distributions with a spread that
is too narrow compared to the true (unknown) distribution, but
the RPS only penalizes bias to the degree that a biased functional
forecast necessarily stems from an uncalibrated density forecast.
See~\citep{Gne11} and~\citep{GK14} for a detailed discussion
on proper scoring rules and their properties.

In continuous time, the actual cumulative probability distribution
of stockout for a SKU is given by
\begin{equation}
F_u(t) = \left\{
\begin{array}{l}
0, \mbox{ for } 0 \leq t \leq u, \\
1, \mbox{ for } u < t \leq d,
\end{array}
\right.
\label{F}
\end{equation}
where $d$ is the number of observation days and $u$ is the day
on which the SKU is actually sold-out.
The forecast cumulative probability distribution $G(t)$
will be evaluated using 
\begin{equation}
\mbox{RPS} = \int_0^d (F_u(t)-G(t))^2 \,dt \,,
\label{RPS}
\end{equation}
and $\left< \mbox{RPS} \right> = E[\mbox{RPS}]$
where the expectation value $E[\dots]$ is calculated over
the distribution of the random variable $U$,
representing the stockout day.

\subsubsection{Uniform distribution of stockout time}
\label{Sec:uni}

To establish a baseline for the metric in our problem of stockout,
we assume a uniform probability density function for $U$,
\begin{equation}
f(u) = \left\{
\begin{array}{l}
1/d, \mbox{ for } 0 \leq u \leq d \,, \\
0, \mbox{ otherwise.}
\end{array}
\right.
\label{unidensity}
\end{equation}

Thus, the cumulative probability function results
\begin{equation}
G_u(t) = t/d, \mbox{ for } 0 \leq t \leq d \,,
\label{uni}
\end{equation}
and we get
\begin{equation}
F_u(t)-G_u(t) = \left\{
\begin{array}{l}
-G_u(t) = -t/d, \mbox{ for } 0 \leq t \leq u \,, \\
1-G_u(t) = 1-t/d, \mbox{ for } u < t \leq d.
\end{array}
\right.
\end{equation}
Therefore,
\begin{equation}
\mbox{RPS}(u) =
\displaystyle\frac{d}{3} \left( 1 +
\frac{3}{d} \left( \frac{u^2}{d}-u \right) \right) \,.
\label{diffunieq}
\end{equation}
On the other hand, using Eq.(\ref{unidensity}), we get
$E[U^n] = \displaystyle\frac{d^n}{n+1}$.
In this way, using Eqs.~(\ref{RPS}) and~(\ref{diffunieq}),
we finally obtain
\begin{equation}
\left< \mbox{RPS} \right> = \displaystyle\frac{d}{6} \,,
\mbox{Var[RPS]} = \displaystyle\frac{d^2}{180} \,.
\label{baseline}
\end{equation}
%

\subsubsection{Forecasting a single day at random}
\label{Sec:oneday}

If the forecast is a single day $u_0$ (point forecast),
the cumulative probability function results
\begin{equation}
G_{u_0}(t) = \left\{
\begin{array}{l}
0, \mbox{ for } 0 \leq t \leq u_0, \\
1, \mbox{ for } u_0 < t \leq d.
\end{array}
\right.
\label{F0}
\end{equation}
Then,
\begin{equation}
\int_0^d (F_u(t)-G_{u_0}(t))^2 \,dt = |u-u_0| \,.
\label{diffdayeq}
\end{equation}
Thus, Eqs.~(\ref{RPS}) and~(\ref{diffdayeq}) we get,
\begin{equation}
\left< \mbox{RPS} \right> = \frac{1}{d} \,\int_0^d |u-u_0| \,du
= \frac{1}{2 \,d} \left( u_0^2 + (d-u_0)^2 \right) \,.
\label{RPSu0}
\end{equation}
The mean RPS under a uniform probability density function for $U$,
has a minimum value when  $u_0=d/2$,
$\left< RPS \right>_{\mbox{min}} = \displaystyle\frac{d}{4}$.
By this way, taking the value of $u_0$ uniformly distributed
in the interval $[0,d]$, from Eq.~(\ref{RPSu0}) we obtain
$\left<\left<\mbox{RPS} \right>\right> = \displaystyle\frac{d}{3}$.
Clearly, this is not the uniformly distributed case given
in Sec.~\ref{Sec:uni}.
We can observe that an somewhat inaccurate but consistent
forecasting approach is preferable to a randomly accurate forecast.
In other words, having a scattershot and inaccurate forecast
yields more value compared to a forecast that is randomly accurate.

\subsection{Evaluation database}
\label{Sec:Data}

Numerous datasets containing sales time series are available
for download. Nevertheless, only a small fraction of these
datasets originate from real-world markets, possess large amounts
of data (tens of thousands of datapoints), have traceable sources,
and are openly accessible with thorough documentation.
Among the valuable datasets, that satisfy the aforementioned
criteria, we identified the dataset released
as part of the Bimbo competition~\citep{bimbo2016}.
This comprehensive dataset comprises 74,180,464 weekly
transactions from Grupo Bimbo S.A.B.~\citep{GRBMF} in Mexico,
featuring 1,799 SKUs and 880,604 unique customers.
Another notable example is the Makridakis competition
M5~\citep{MSA22a}, which focuses on retail time series
using real-life data from Walmart Inc.~\citep{WMT}.
It involves 3,049 SKUs and 42,840 hierarchical daily time series.

MercadoLibre~Inc.~\citep{meli} is an Argentine company founded
in 1999 that operates the largest e-commerce marketplace
in Latin America, connecting a network of more than 140 million
active users and over 1 million active sellers across
the 18 countries where the company operates.
Between July and September 2021, the company hosted
a MELI Data Challenge, which tasked participants
with forecasting the stockout day for the inventory of given SKUs.
For this challenge, a vast database was released~\citep{meli2021}.
This dataset comprises daily sales time series for 660,916
distinct SKUs, corresponding to the months of February and March 2021.
The time series exhibit a wide variety of sales behaviors,
ranging from slow-moving items to highly demanded inventories.
Many series in the competition exhibit intermittency,
characterized by sporadic unit sales that result in numerous
days with zero sales.
Furthermore, not all SKUs were offered for sale every day
of the month.

Additionally, for each SKU, supplementary data on the item's
characteristics and product family it belongs were provided
in a separate database.

The task involved predicting the probability $P(0,k)$ for each SKU,
using training data from February and March,
given an initial stock of $m$ items at the beginning of April,
where $k$ represents the days in April 2021.
The website hosting the challenge is no longer available online.
However, the competition leaderboard indicated that the mean
RPS across all SKUs was $3.71$ for the participant ranked first.
We adopt this value as a benchmark for evaluating our methodological
framework. Dispersion metrics for RPS values were not provided
in the leaderboard.

The test dataset containing April's data was never released.
This constraint forces us to retain only the February data
for training and use the March data for testing.
Due to this limitation, despite the database containing
660,916 distinct SKUs, we find that only 495,353 of them
have sales data for both months.

As an example, in Table~\ref{SKU_538100}, we present the sales data
for SKU 538100, a smartphone accessory, over a 28-day period
in February (top line) and a 31-day period in March (second line).
For each day with sales in March, we conducted an evaluation.
We evaluated the stockout behavior of SKU 538100 by considering
different hypothetical initial stock levels at the beginning
of March.
Specifically, we used the cumulative sales up to each day
in March as a hypothetical initial stock level and determined
the corresponding stockout time.
The results are shown in Table~\ref{SKU_538100}, where the values
of $m$, given in the bottom line, represent the initial stock
levels aligned with the corresponding stockout day.
This approach allows us to perform evaluations for each SKU
with sales in both months, with the number of evaluations equal
to the number of days with sales in March.
\begin{table}[ht]
\renewcommand{\arraystretch}{0.7}
\setlength{\tabcolsep}{3pt}
\begin{tabular}{cccccccccccccccccccccccccccccccc}
sales &
0 & 0 & 2 & 1 & 2 & 0 & 0 & 0 & 0 & 1 & 0 & 2 & 1 & 0 & 0 & 0 & 0 & 0 & 0 & 1 & 0 & 0 & 2 & 1 & 0 & 0 & 1 & 1 \\
\\
sales &
0 & 1 & 2 & 0 & 0 & 0 & 1 & 0 & 1 & 0 & 3 & 1 & 0 & 0 & 1 & 0 & 1 & 1 & 0 & 0 & 0 & 0 & 0 & 2 & 0 & 1 & 0 & 1 & 2 & 3 & 4 \\
$m$ &
  & 1 & 3 &   &   &   & 4 &   & 5 &   & 8 & 9 &   &   &10 &   &11 &12 &   &   &   &   &   &14 &   &15 &   &16 &18 &21 &25
\end{tabular}
\caption{Sales of SKU 538100 in February 2021 (top line)
and during March 2021 (middle line). The bottom line displays
the possible initial stocks, $m$, in the places corresponding
to the stockout times in March.}
\label{SKU_538100}
\end{table}

\begin{figure}[!th]
\begin{center}
\includegraphics[width=0.90\textwidth]{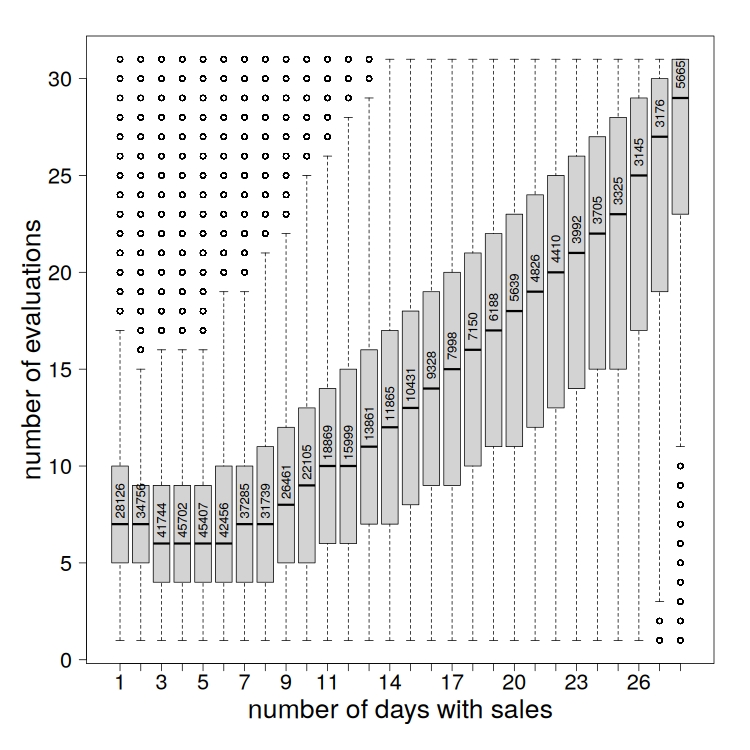}
\caption{Boxplots of the total number of feasible evaluations
in March based on days with sales in February.
The numbers inside the boxes indicate the number
of SKUs with that number of sales in February.
Circles represent outliers.}
\label{evaluations}
\end{center}
\end{figure}

Thus, using this augmentation method, we construct a test set
comprising 4,822,218 distinct pairs of values ($m$, stockout day),
derived from the March time series data involving 495,353 different
SKUs.
In Figure~\ref{evaluations}, the boxplots depict the number
of evaluation pairs, given the number of days with sales in February.
%

\subsection{Naive frequentist approach}
\label{Sec:Freq}

The naive frequentist model for demand presented in
Sec.~\ref{Sec:naive}, although not the most advanced method,
remains a widely adopted and practical option for many inventory
management applications, despite being susceptible to statistical
flukes.
The values of  $\alpha_\ell$ corresponding to this model
are calculated by determining the frequencies of sales
in the training set, which consists of data from February,
as shown in Table~\ref{alpha_538100} for SKU 538100.
\begin{table}[ht]
\begin{center}
\begin{tabular}{crc}
sales & freq. & $\alpha$ \\
0 & 17 & 0.61 \\
1 &  7 & 0.25 \\
2 &  4 & 0.14
\end{tabular}
\end{center}
\caption{Values of $\alpha$ corresponding to SKU 538100.}
\label{alpha_538100}
\end{table}
To make predictions for initial stocks $m>2$,
the table of frequencies is complemented with
$\alpha_\ell = 0$ for all $2 < l \leq m$.
For high values of $l$, $\alpha_\ell$ may be underestimated
due to the electronic marketplace's restrictions on purchases
exceeding current stock.
Thus, we cannot determine the true interest of shoppers
when they want to buy a large number of items of a given SKU
with limited stock.

After calculating $\alpha_\ell$ for each SKU with the procedure
indicated above, we can evaluate our stochastic model complemented
with the naive frequentist demand approach, using the
algorithm~(\ref{algo}) to construct $P(0,k)$ corresponding
to each evaluating datapoint in the forecasting database.
We will refer to it as the NFQ model.
Given that we are forecasting over a finite interval of $d$ days,
it is convenient to normalize previously the stockout probability
by considering the ratio $G(k) = P(0,k)/P(0,d)$,
where $0 \leq k \leq d$.
Finally we measure the resulting RPS for each evaluation according
to the discrete version of Eq.(\ref{RPS}),
\begin{equation}
\mbox{RPS} = \sum_{k=1}^d (F_u(k)-G(k))^2 \,,
\label{RPSd}
\end{equation}
where $F_u(k)$ is given by Eq.(\ref{F}), $u$ is equal to the
stockout day of the inventory, and $d=31$ for March.

\subsection{Parameters estimation}
\label{Sec:parameters}

The parametric distributions for demand discussed
in Sec.~\ref{Sec:Demand} can be implemented by the method
of moments~\citep{JKK05} for parameter estimation.
However, no single distribution is universally
applicable across all SKUs.

The Poisson distribution is based on the assumption
of independently random arriving at a uniform rate.
The mean ($\mu$) and variance ($\sigma^2$)
are $\mu=\sigma^2=\lambda$.
When the rate of arrivals is non-uniform, the Binomial and
Negative Binomial distributions are more appropriate
for describing the demand.
The mean and variance of the Binomial distribution are
\begin{equation}
\mu =  C \,p\,, \; \sigma^2 = C \,p \,(1-p) \,,
\end{equation}
whereas for the Negative Binomial distribution are
\begin{equation}
\mu = r \frac{1-p}{p} \\, \; \sigma^2 =r \frac{1-p}{p^2} \,.
\end{equation}
Thus, these distributions permit independent selection
of the mean and variance; however, whereas the Binomial
distribution is suitable when $\sigma^2 < \mu$,
the Negative Binomial distribution applies in the opposite case.
In contrast, the Poisson case does not have this restriction.

The methods of moments~\citep{JKK05} allows fitting
the parameters of our interest from the maximum likelihood
estimators for mean (sample average $\bar{x}$) and variance
(sample variance $s^2$) of demand and according to
\begin{equation}
\begin{array}{lll}
\lambda = \bar{x} &
& \mbox{(Poisson)} \\
p = 1 - \displaystyle\frac{s^2}{\bar{x}} \,,
& C = \displaystyle\frac{\bar{x}^2}{\bar{x}-s^2}
& \mbox{(Binomial)} \\
p = \displaystyle\frac{\bar{x}}{s^2} \,,
& r = \displaystyle\frac{\bar{x}^2}{s^2-\bar{x}} \,.
& \mbox{(Negative Binomial)}
\end{array}
\label{parameters}
\end{equation}
Therefore, for parameter estimation, we need to calculate
the average ($\bar{x}$) and sample variance ($s^2$) of sales
that occurred in February.
According to Eq.~(\ref{parameters}), we can use the Binomial
distribution to model demand if $\bar{x} > s^2$ and
the Negative Binomial distribution if $s^2 > \bar{x}$.
When using these distribution to model demand,
we encounter a limitation due to the fact that
the upper incomplete gamma function,
the regularized incomplete beta function, and factorial
make numerical evaluation impractical for large values of $m$.

In this manner, to make forecasts in all cases,
we will use Eqs.~(\ref{BinomialP0}),
(\ref{NBinomialP0}), and (\ref{PoissonP0})
This model implements either the Binomial or Negative Binomial
distribution to model $\alpha_\ell$, based on the relationship
between $\bar{x}$ and $s^2$, while reserving the Poisson
distribution only for the special case where $\bar{x} = s^2$.
We will refer to this hybrid model of stockout as the BNBP model.

To calculate RPS, we use Eq.~(\ref{RPSd}) again, with
$G(k) = P(0,k)/P(0,d)$, where $1 \leq k \leq d=31$.

\section{Forecasting results}
\label{sec:FR}

In Table~\ref{summary}, we present a statistical summary
of the forecasting performed with the
BNBP (defined in Sec.~\ref{Sec:parameters}) and
NFQ (defined in Sec.~\ref{Sec:Freq}) models over
the evaluation set for stockouts constructed
in Sec.~\ref{Sec:Data}.
For comparison, we also include the case in which we complement
our stochastic model for stockouts using only the Poisson
distribution for demand.
\begin{table}[htbp]
\begin{center}
\begin{tabular}{l|cccccc}
model & min & 1st Q & median & mean & 3rd Q & max \\
\hline
Poisson & 0.00 & 1.50 & 3.44 & {\bf 5.32} & 7.73 & 31.00 \\
NFQ     & 0.00 & 1.57 & 3.18 & {\bf 4.91} & 6.93 & 31.00 \\
BNBP    & 0.00 & 1.56 & 3.11 & {\bf 4.78} & 6.70 & 31.00 \\
\end{tabular}
\end{center}
\caption{Statistical summary of RPSs obtained in the evaluations
for the different models.}
\label{summary}
\end{table}

Table~\ref{full} collects the mean and standard deviation
of RPSs generated by all evaluations of our stochastic forecasting
scheme with the different demand distributions.
For comparison, it includes the baseline value corresponding
to the random uniform distribution for stockout time given
by Eq.~(\ref{baseline}) with $d=31$ and the benchmark value
mentioned in Sec.~\ref{Sec:Data}.
\begin{table}[ht]
\begin{minipage}{\textwidth}
\begin{center}
\begin{tabular}{l|rrrrcc}
\hline
model & SKUs & & evals. & & mean & sd \\
\hline
Baseline  &&&&                      & {\bf 5.2} & 2.3 \\
Poisson   & 495,164 & & 4,653,044 & & {\bf 5.3} & 5.1 \\
NFQ       & 495,353 & & 4,822,218 & & {\bf 4.9} & 4.6 \\
BNBP      & 495,352 & & 4,821,348 & & {\bf 4.8} & 4.4 \\
Benchmark &&&&                      & {\bf 3.7} & NA  \\
\hline
\hline
BNBP & SKUs
& $\%$\footnote{Percentage over 495,352 SKUs.}
& evals.
& $\%$\footnote{Percentage over 4,821,348 evaluations.}
& mean & sd \\
\hline
P  &  27,308 & $ 5.5$ &   204,100 & $ 4.2$ & {\bf 7.1} & 5.7 \\
NB & 352,608 & $71.2$ & 3,709,790 & $77.0$ & {\bf 4.6} & 4.3 \\
B  & 115,436 & $23.3$ &   907,458 & $18.8$ & {\bf 5.0} & 4.6 \\
\hline
\end{tabular}
\caption{Mean and standard deviation of RPSs generated
by different models, alongside the corresponding number
of SKUs and evaluation performed for each of them.
The lower panel shows the isolated values for the BNBP components:
Poisson (P), Negative Binomial (NB), and Binomial (B) demand
distributions.
}
\label{full}
\end{center}
\end{minipage}
\end{table}
Additionally, this table breaks down the BNBP values
to examine the performance of each constituent model.
Most BNBP evaluations are performed with the Negative Binomial
distribution because the variance in the number of sales exceeds
the mean in most cases.

At this point, it's worth noting that in the MELI competition,
from where we extract the benchmark value, participants were required
to forecast a single probability distribution, $G(k)$, per SKU and $m$
for April, using February and March data to train their machine
learning models and leveraging all available contextual information
about the SKUs.
Notably, the training sales series in the competition were twice
as long as those used for setting up our model.

\begin{figure}[!th]
\begin{center}
\includegraphics[width=0.90\textwidth]{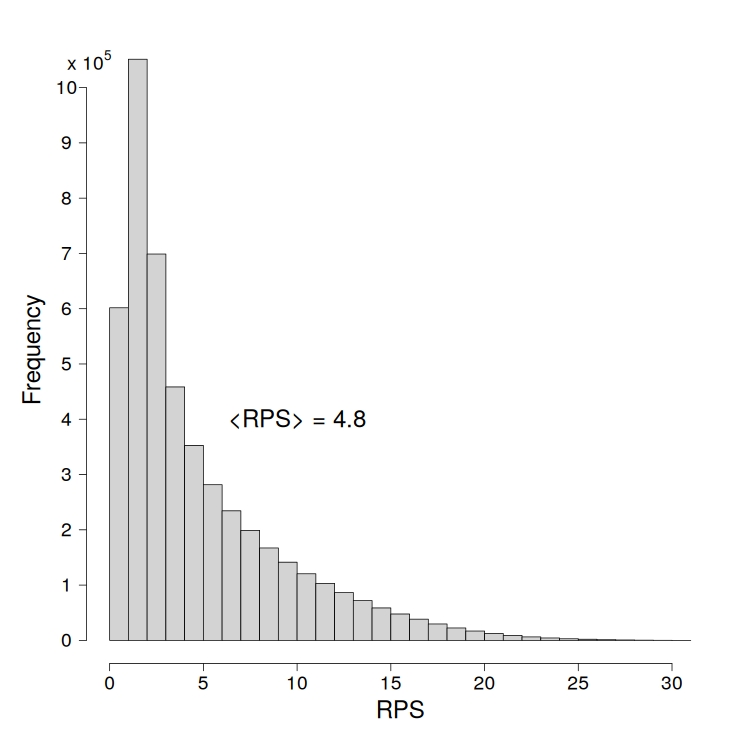}
\caption{Histogram of RPS values from evaluations using
the BNBP model.}
\label{BNBPhist}
\end{center}
\end{figure}

Figure~\ref{BNBPhist} shows the histogram of RPS values
obtained with the BNBP model. The global mean value,
$\left< \mbox{RPS} \right> = 4.8$, is skewed upward over
the median, as shown in Table~\ref{summary}, by the values
in the tail observed in this figure.

\begin{figure}[!th]
\begin{center}
\includegraphics[width=0.90\textwidth]{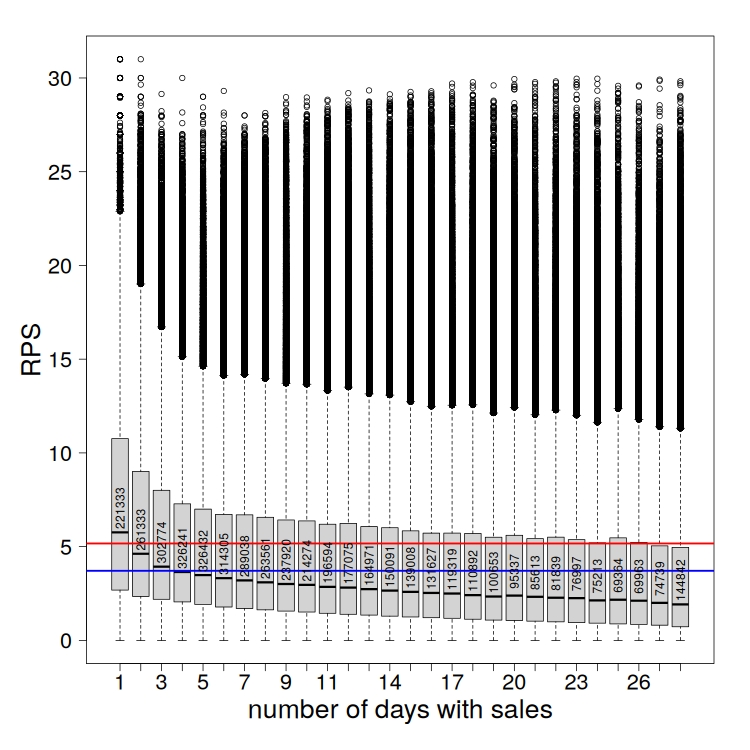}
\caption{Boxplots of RPSs from BNBP model evaluations versus
days with sales in February (training set).
The number of evaluations is indicated within each box.
The red line indicates the baseline value, and
the blue line represents the benchmark.
Outliers are represented by circles.}
\label{BNBPrps}
\end{center}
\end{figure}

\begin{figure}[!th]
\begin{center}
\includegraphics[width=0.90\textwidth]{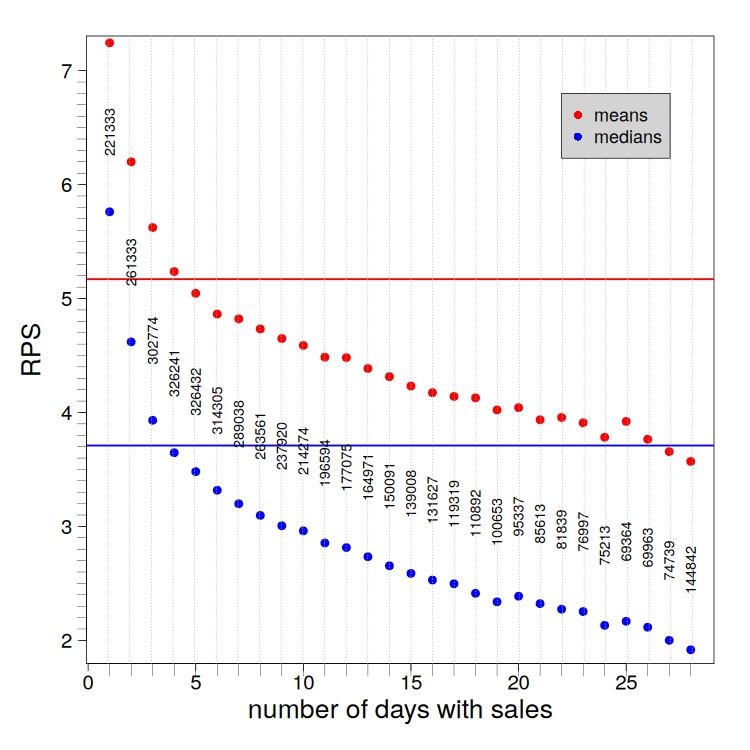}
\caption{Medians and means of RPSs from BNBP model evaluations
versus days with sales in February (training set).
The numbers of evaluations are indicated aligned
with each pair of dots.
The red line indicates the baseline value, and
the blue line represents the benchmark.}
\label{BNBPm&m}
\end{center}
\end{figure}

For a better visualization of the results, Figure~\ref{BNBPrps}
presents boxplots of RPSs obtained from evaluating the BNBP model,
stratified by the number of days with sales in February.
The horizontal red line indicates the baseline value of $5.17$,
derived from the random uniform distribution given
by Eq.~(\ref{baseline}) with $d=31$.
In contrast, the blue line represents the benchmark value of $3.71$,
achieved by the winner of the MELI data challenge.
The distributions of RPSs are skewed, and the quartiles decrease
monotonically with the number of days with sales in February.
In Figure~\ref{BNBPm&m}, the region between the baseline
and benchmark values is zoomed in, showing only the medians
and means of the RPSs for the BNBP model evaluations.
When SKUs have sales on more than 3 days in February,
the median RPS is consistently below the benchmark,
whereas when SKUs have sales on more than 4 days
in the training set, the mean RPS is consistently below
the baseline.

The NFQ model evaluations show similar behavior, as depicted
in Figures~\ref{BNBPhist}, \ref{BNBPrps}, and~\ref{BNBPm&m}.

By construction, we know that all stockouts in the evaluation
dataset occur in March. However, in some datapoints,
the model's non-normalized prediction on the last day falls
below $0.5$ ($P(0,d) < 0.5$, $d=31$).
In such cases, our model erroneously predicts stockouts
in the next month, indicating it is unable to make reliable
predictions within March. We can exclude these specific
evaluations from our analysis.
With this criterion, we can analyze only $67\%$ of the total
universe of 4,822,218 possible evaluations, involving about
$98\%$ of SKUs in the database for the NFQ and BNBP models.
Table~\ref{thr=0} shows the mean and standard deviation
of RPSs filtered according to this exclusion criterion.
A consistent improvement is observed in all models.
\begin{table}[ht]
\begin{minipage}{\textwidth}
\begin{center}
\begin{tabular}{l|rrrrcc}
\hline
model & SKUs
& $\%$\footnote{Percentage over 495,353 SKUs with data in both months.}
& evals.
& $\%$\footnote{Percentage over 4,822,218 possible evaluations.}
& mean & sd \\
\hline
Baseline &&&&                                    & {\bf 5.2} & 2.3 \\
Poisson  & 485,515 & $98.0$ & 3,179,806 & $65.9$ & {\bf 4.5} & 4.4 \\
NFQ      & 485,593 & $98.0$ & 3,254,422 & $67.5$ & {\bf 4.2} & 3.9 \\
BNBP     & 485,440 & $98.0$ & 3,242,625 & $67.2$ & {\bf 4.2} & 3.9 \\
Benchmark   &&&&                                 & {\bf 3.7} & NA  \\
\hline
\hline
BNBP & SKUs
& $\%$\footnote{Percentage over 485,440 SKUs.}
& evals.
& $\%$\footnote{Percentage over 3,242,625 evaluations.}
& mean & sd \\
\hline
P  &  22,185 & $ 4.6$ &    51,456 & $ 1.6$ & {\bf 4.5} & 3.3 \\
NB & 349,753 & $72.0$ & 2,650,021 & $81.7$ & {\bf 4.3} & 4.0 \\
B  & 113,502 & $23.4$ &   541,148 & $16.7$ & {\bf 4.1} & 3.6 \\
\hline
\end{tabular}
\caption{Mean and standard deviation of filtered RPSs generated
by different models, alongside the corresponding number
of SKUs and evaluation counts performed for each of them.
The lower panel shows the isolated values for the BNBP components:
Poisson (P), Negative Binomial (NB), and Binomial (B) demand
distributions.
}
\label{thr=0}
\end{center}
\end{minipage}
\end{table}

Summarizing our results, assuming a Poisson distribution for demand
implies lower performance in our forecasting model measured by RPS
compared to the other demand distributions.
As a rule of thumb, we find that modeling demand with the
Binomial distribution in our model generates many predictions
outside March. However, when we apply the exclusion criterion,
in cases where the Binomial demand is applicable
(in the training series $s^2 < \bar{x}$), our stochastic model
achieves the best performance.
The NFQ approach is a reasonable option in all cases,
yielding results comparable to the BNBP alternative.

\section{Concluding remarks}
\label{sec:fin}

We have developed a stochastic model for the distribution
of stock over time for inventories without replenishment.
Our general framework requires only the initial stock
and the known series of sales of the specific selected SKU
as input. As an added benefit, we have calculated the probability
of frustrated sales due to stockouts.
For specific analytical demand distributions (deterministic,
Poisson, binomial, and negative binomial), we derive closed-form
expressions for the time-dependent stock distribution.

Our results enable the prediction of stockouts, and we evaluated
the reliability of our predictions using a massive dataset
of real-world examples from an electronic marketplace,
covering SKUs with different behaviors and achieving robust
performance even with very short historical sales data
for short planning horizons.

The choice of model may hinge on the specific priorities
of the application (e.g., coverage, accuracy, implementation costs).
Machine learning models could perform better than simpler
statistical models, but forecasting for a single SKU
necessitates training on the complete database with all
SKUs~\citep{BR18,HMF19,Pav19}.
In contrast, our model only requires the sales time series
of the SKU for which we want to make forecasts.
It then evaluates a closed-form expression for analytical
demand distributions or implements Algorithm~\ref{algo}
for the naive frequentist approach to demand.
This results in significantly lower computational effort.
The cost reduction has a substantial impact on future
maintenance, particularly with additional data ingestion.
This characteristic makes our model particularly appealing
to inventory managers and OR practitioners, as it enables
efficient forecasting and decision-making processes,
even in large-scale operational settings.

SKU is the lowest item descriptor in a marketplace.
Our stochastic model disregards correlations among them,
such as the cross-selling effect~\citep{ZZZ+14,LLFH23}.
For instance, stockouts of a given SKU could potentially
impact sales of another SKU for a closely related product,
like the same item in a different size or color,
or from a different brand.
These nuances could be captured by top-performing machine
learning models, such as tree-based models like
XGBoost~\citep{Nie16}.
However, our simpler model enhances interpretability
and facilitates causal inference, enabling us to isolate
the effects of demand on a single SKU.
In this way, our contribution lays the groundwork for future
research on how correlations among SKUs affect stockout forecasting.


\vspace{0.2in}
\subsection*{Acknowledgments}
This research received no external funding.

\subsection*{Declaration of competing interest}
The author declares that he has no known competing financial
interests or personal relationships that could have appeared
to influence the work reported in this article.

\subsection*{Data and code availability}
The data used in the analyses are publicly available
and can be accessed through the source cited in~\citet{meli2021}.
In addition, the R code used for the experiments is available
upon request.

\vspace{0.1in}
\appendix
\section{Appendix}
\label{Appendix}
\numberwithin{equation}{section}

Eqs.~(\ref{deterPn}), (\ref{PoissonPn}), (\ref{BinomialPn}),
and (\ref{NBinomialPn}) can be proven by complete induction.
In all four situations, the base case is immediate,
$P(n,k=1) = \alpha_{m-n}$.

For the deterministic case, using Eq.~(\ref{deltas}),
it is inmediate to obtain,
\begin{equation}
\begin{array}{ll}
\displaystyle\sum_{\ell=n}^m \alpha_{\ell-n} P(\ell,k-1)
& =
\displaystyle\sum_{\ell=n}^m
\delta_{\ell-n,h} \,\delta_{\ell,m-(k-1)h}
\\
& = \delta_{n+h,m-(k-1)h}
\\
& = \delta_{n,m-kh} \,.
\end{array}
\end{equation}
In the context of the Poisson model, the induction step
can be accomplished using the binomial expansion~(\ref{binomial}).
Thus, we obtain
\begin{equation}
\begin{array}{ll}
\displaystyle\sum_{\ell=n}^m \alpha_{\ell-n} P(\ell,k-1)
& =
e^{-k \,\lambda} \displaystyle\sum_{\ell=n}^m
\displaystyle\frac{\lambda^{\ell-n}}{(\ell-n)!}
\displaystyle\frac{((k-1) \,\lambda)^{m-\ell}}{(m-\ell)!}
\\
& =
e^{-k \,\lambda} \,\lambda^{m-n} \displaystyle\sum_{j=0}^{\bf m-n}
\displaystyle\frac{(k-1)^{{\bf m-n}-j}}{j! \,({\bf m-n} - j)!}
\\
& =
e^{-k \,\lambda} \,\lambda^{m-n}
\displaystyle\frac{k^{m-n}}{(m-n)!} \,.
\end{array}
\end{equation}
For the Binomial model, by using Vandermonde
identity~(\ref{Vandermonde}) we obtain
\begin{equation}
\begin{array}{ll}
\displaystyle\sum_{\ell=n}^m \alpha_{\ell-n} P(\ell,k-1)
& =
\displaystyle\sum_{\ell=n}^m
\binom{C}{\ell-n} \, p^\ell \,q^{C-\ell+n} \,
\binom{(k-1)C}{m-\ell} \,p^{m-\ell} \,q^{(k-1)C-m+\ell}
\\
& =
\displaystyle\sum_{j=0}^{\bf m-n}
\binom{C}{j} \,\binom{(k-1)C}{{\bf m-n}-j} \,
p^{m-n} \,q^{kC-(m-n)}
\\
& =
\displaystyle\binom{kC}{m-n} \,p^{m-n} \,q^{kC-(m-n)} \,.
\end{array}
\end{equation}
Last, for the Negative Binomial model by using Vandermonde
convolution~(\ref{Vandermonde2}) we obtain
\begin{equation}
\begin{array}{ll}
\displaystyle\sum_{\ell=n}^m \alpha_{\ell-n} P(\ell,k-1)
& =
\displaystyle\sum_{\ell=n}^m
\binom{r-1+\ell-n}{\ell-n} \, p^r \,q^{\ell-n} \,
\binom{(k-1)r-1+m-\ell}{m-\ell} \,p^{(k-1)r} \,q^{m-\ell}
\\
& =
\displaystyle\sum_{j=0}^{\bf m-n}
\binom{r-1+j}{j} \,\binom{(k-1)r-1+{\bf m-n}-j}{{\bf m-n}-j} \,
p^{kr} \,q^{m-n}
\\
& =
\displaystyle\binom{kr-1+m-n}{m-n} \,p^{kr} \,q^{m-n} \,.
\end{array}
\end{equation}
By using the last four equations, we can reproduce the recurrence
Eq.~(\ref{MEn}).


\bibliographystyle{apalike}
\bibliography{newsv_apa}




\end{document}